\documentclass[reprint, superscriptaddress, secnumarabic, amssymb, nobibnotes, aps, prl]{revtex4-2}

\pdfoutput=1

\usepackage{tikz}
\usetikzlibrary{positioning}
\usepackage{graphicx}
\usepackage{epstopdf}
\usepackage[T1]{fontenc}
\usepackage[utf8]{inputenc}
\DeclareUnicodeCharacter{2212}{~}
\DeclareUnicodeCharacter{0327}{~}
\usepackage[utf8]{inputenc}
\usepackage{amsbsy}
\usepackage{gensymb}
\usepackage[T1]{fontenc}
\usepackage{amsmath}
\usepackage{amssymb}
\usepackage{bbm}
\usepackage{physics}
\usepackage{braket}
\usepackage{xcolor}
\usepackage{comment}
\usepackage{float}
\allowdisplaybreaks
\usepackage{graphicx}
\usepackage[colorlinks=true]{hyperref}  
\hypersetup{
    bookmarks=true,         
    unicode=false,          
    pdftoolbar=true,        
    pdfmenubar=true,        
    pdffitwindow=false,     
    pdfstartview={FitH},    
    pdftitle={Evidence for conventional superconductivity in Bi$_2$PdPt and prediction of topological superconductivity in disorder-free $\gamma$-BiPd},    
    pdfauthor={S. Sharma}, 
    pdfsubject={},   
    pdfcreator={},   
    pdfproducer={}, 
    pdfkeywords={} {} {}, 
    pdfnewwindow=true,      
    colorlinks=true,       
    linkcolor=blue, 
    citecolor=blue,        
    filecolor=magenta,      
    urlcolor=blue           
} 
\usepackage[normalem]{ulem}
\usepackage{orcidlink}

\renewcommand{\approx}{\simeq}

\renewcommand{\vec}[1]{\boldsymbol{#1}}

\definecolor{lred}{HTML}{EB212E}

\begin{document}
    
\preprint{APS/123-QED}

\title{Evidence for conventional superconductivity in Bi$_2$PdPt and prediction of topological superconductivity in disorder-free $\gamma$-BiPd  }

\author{S.~Sharma\,\orcidlink{0000-0002-4710-9615}}
\affiliation{Department of Physics and Astronomy, McMaster University, Hamilton, Ontario L8S 4M1, Canada}
\author{A. D. S. Richards\,\orcidlink{0000-0003-4919-104X}}
\affiliation{Department of Physics and Astronomy, McMaster University, Hamilton, Ontario L8S 4M1, Canada}
\author{Sajilesh K. P.\,\orcidlink{0000-0002-9921-4062}}
\affiliation{Physics Department, Technion-Israel Institute of Technology, Haifa 32000, Israel}
\author{A. Kataria}
\affiliation{Department of Physics, Indian Institute of Science Education and Research Bhopal, Bhopal, 462066, India}
\author{B. S. Agboola}
\affiliation{Department of Materials Science and Engineering, McMaster University, 1280 Main St W, Hamilton, L8S 4L8, ON, Canada}
\author{M.~Pula\,\orcidlink{0000-0002-4567-5402}}
\affiliation{Department of Physics and Astronomy, McMaster University, Hamilton, Ontario L8S 4M1, Canada}
\author{J. Gautreau}
\affiliation{Department of Physics and Astronomy, McMaster University, Hamilton, Ontario L8S 4M1, Canada}
\author{A. Ghara}
\affiliation{Indian Institute of Science Education and Research Pune, Pune, 411008, India}
\author{D. Singh}
\affiliation{Department of Physics, Indian Institute of Science Education and Research Bhopal, Bhopal, 462066, India}
\author{S. Marik}
\email[]{Current address: Department of Physics and Materials Science, Thapar Institute of Engineering and Technology, Patiala, Punjab 147004}
\affiliation{Department of Physics, Indian Institute of Science Education and Research Bhopal, Bhopal, 462066, India}
\author{S. R. Dunsiger}
\affiliation{TRIUMF, Vancouver, British Columbia V6T 2A3, Canada}
\author{M. J. Lagos}
\affiliation{Department of Materials Science and Engineering, McMaster University, 1280 Main St W, Hamilton, L8S 4L8, ON, Canada}
\affiliation{Canadian Centre for Electron Microscopy, McMaster University, 1280 Main St W, Hamilton, L8S 4M1, ON, Canada}
\author{
A. Kanigel}
\affiliation{Physics Department, Technion-Israel Institute of Technology, Haifa 32000, Israel}
\author{E.~S. S{\o}rensen\,\orcidlink{0000-0002-5956-1190}}
\affiliation{Department of Physics and Astronomy, McMaster University, Hamilton, Ontario L8S 4M1, Canada}
\author{R.~P.~Singh\,\orcidlink{0000-0003-2548-231X}}
\affiliation{Department of Physics, Indian Institute of Science Education and Research Bhopal, Bhopal, 462066, India}
\author{G.~M.~Luke\,\orcidlink{0000-0003-4762-1173}}
\email[]{luke@mcmaster.ca}
\affiliation{Department of Physics and Astronomy, McMaster University, Hamilton, Ontario L8S 4M1, Canada}
\affiliation{TRIUMF, Vancouver, British Columbia V6T 2A3, Canada}

\date{\today}

\begin{abstract}
\begin{flushleft}
\end{flushleft}
We present comprehensive investigations into the structural, superconducting, and topological properties of Bi$_2$PdPt. 
Magnetization and heat capacity measurements performed on polycrystalline Bi$_2$PdPt  demonstrate a superconducting transition at $\approx$ 0.8 K. Moreover, muon spin relaxation/rotation ($\mu$SR) measurements present evidence for a time reversal symmetry preserving, isotropically gapped superconducting state in Bi$_2$PdPt. We have also performed density-functional theory (DFT) calculations on Bi$_2$PdPt alongside the more general isostructural systems,  BiPd$_{x}$Pt$_{1-x}$, of which Bi$_2$PdPt and $\gamma$-BiPd are special cases for $x=0.5$ and $x=1$ respectively.
We have calculated the $Z_2$ topological index from our DFT calculations for a range of substitution fractions, $x$, between $x=0$ and $x=1$ characterizing the topology of the band structure. 
We find a non-trivial topological state when $x>0.75$ and a trivial topological state when $x<0.75$. Therefore our results indicate that  BiPd$_{x}$Pt$_{1-x}$ could be a topological superconductor for $x>0.75$.
\end{abstract}
\maketitle
    
\section{\label{sec:level1}INTRODUCTION \protect\\ }
Bi- and Pd-based materials have attracted interest recently as possible realizations of topological superconductivity~\cite{sun2015dirac, Benia_Wahl_2016, Dimitri_Neupane_2018, Sakano2015, Lv_Xue_2017}. The relatively high spin-orbit coupling in these materials can lead to a continuous, direct band gap and band inversion, allowing for non-trivial band topology. For example, Dirac surface states observed in $\alpha$-BiPd \cite{sun2015dirac,Benia_Wahl_2016}, $\alpha$-Bi$_2$Pd \cite{Dimitri_Neupane_2018}, and $\beta$-Bi$_2$Pd \cite{Sakano2015} are attributed to non-trivial band topology. Fu and Kane showed that Dirac surface states in proximity to bulk superconductivity might realize Majorana zero modes~\cite{Fu_Kane_2008}, which may be useful in implementing stable qubits in quantum information devices. This mechanism could explain the apparent realization of Majorana zero modes in thin films of $\beta$-Bi$_2$Pd~\cite{Lv_Xue_2017}.

The apparent non-trivial topological properties and superconductivity observed in other Bi- and Pd-based materials have motivated us to study Bi$_2$PdPt. Bi$_2$PdPt is a particular realization of the series of disordered compounds BiPd$_{x}$Pt$_{1-x}$ for $x=0.5$, in which Pd and Pt atoms occupy the same site with equal probability. Bi$_2$PdPt is isostructural to the $\gamma$ phase of BiPd. In principle, the concentration of Pd and Pt may be tuned to control the strength of spin-orbit coupling and electron-phonon coupling simultaneously.  

We have studied the structural and superconducting properties of Bi$_2$PdPt using microscopic and bulk techniques, including powder X-ray diffraction (XRD), High angle annular dark field  Scanning transmission electron microscopy (HAADF STEM) Imaging,  Energy Dispersive X-ray Spectroscopy (EDS), and muon spin relaxation/rotation ($\mu$SR). 
We find that Bi$_2$PdPt is a 0.8 K bulk superconductor, in contrast to earlier reports, which have suggested a superconducting transition temperature of 4 K for Bi$_2$PdPt \cite{anshu}.  Kataria et al. \cite{anshu} could not observe the 0.8 K transition as they did not measure below 1.8 K. The higher $T_c$ superconductivity  could be attributed to the surface superconductivity. 

To complement our experimental results, we have also performed density functional theory (DFT) calculations to characterize the topological properties. In addition to our analysis of Bi$_2$PdPt, we have explored the possibility of realizing a topological superconductor in the isostructural $\gamma$ phase of BiPd, which, to our knowledge, has not been stabilized in bulk form at low enough temperatures for a transition to a superconducting state to occur. From our theoretical analysis of our DFT results, we find that Bi$_2$PdPt is topologically trivial, while $\gamma$-BiPd is topologically non-trivial. Our results indicate that increasing the concentration, $x$, of Pd in BiPd$_x$Pt$_{1-x}$ from $x=0.5$ (Bi$_2$PdPt) could lead to a topological phase transition at a critical concentration $x_c$ that we have estimated to be near $x_c=0.75$. We find Dirac states in the calculated surface-state spectral function for $\gamma$-BiPd with a binding energy of less than 0.1 eV.




\begin{figure}[]
\begin{center}
\includegraphics[width=0.47\textwidth]{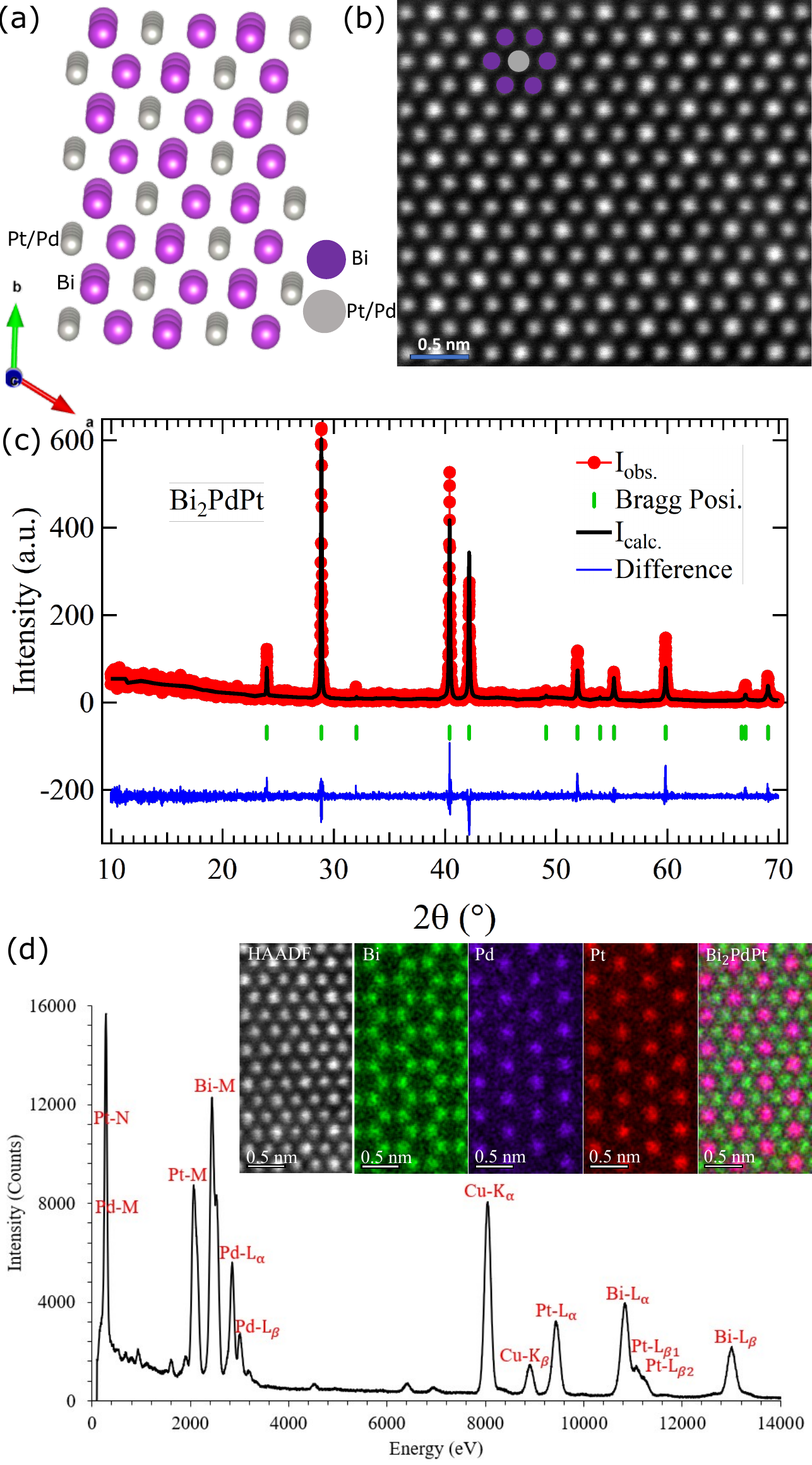}
\caption{\label{XRD} (a) Arrangement of Bi and Pt/Pd columns in the crystal structure of Bi$_2$PdPt when looked along the c-axis. (b) HAADF-STEM image of Bi$_2$PdPt acquired along the c-axis revealing the atomic positions.  (c) The observed powder X-ray diffraction (XRD) pattern and its Rietveld refinement confirm that Bi$_2$PdPt crystallizes in a hexagonal crystal structure with space group $P6_3/mmc$  (d) EDS spectrum of Bi$_2$PdPt showing different x-ray line peaks. [Inset]  Spatially resolved EDS maps acquired with X-ray lines of Bi (green), Pd (purple), Pt (red), and an overlay of Pt and Pd (Magneta) showing the spatial distribution of the atomic columns of the elements in the crystal. Notice that Pd and Pt occupy sites on the same atomic lattice. }
\end{center}
\end{figure}

\section{EXPERIMENTAL DETAILS}
The polycrystalline sample of Bi$_2$PdPt, used in this experiment, was synthesized following the process described in reference \cite{anshu}.  This sample was transferred to a conical quartz tube and lowered across a temperature gradient in a molten, in an effort to obtain a single crystal. We were able to select millimeter-sized crystals which were used for electron microscopy measurements. All the other measurements reported in this manuscript were performed on polycrystalline samples.

The phase purity of the samples was confirmed using powder X-ray diffraction using a PANalytical Xpert$^3$  powder diffractometer. A millimeter-sized crystal obtained from the sample was aligned using Laue diffraction and analyzed using High angle annular dark field  Scanning transmission electron microscopy (HAADF STEM) imaging and  Energy Dispersive X-ray Spectroscopy (EDS). The HAADF-STEM image was acquired on a Spectra Ultra STEM microscope operated at 300kV with an electron probe of convergence semi-angle of 28 mrad, delivering about 0.5 \AA ~ resolution. The STEM images were acquired in HAADF mode. The transmitted high-angle scattered electrons were collected using an annular detector with inner and outer collection angles of 49 mrad and 200 mrad, respectively. The electron beam current for the image acquisition was 84 pA.  The EDS work was conducted at 200 kV using an Ultra-X ray detector, allowing solid angle collections of about 4.45 sr. Spatially-resolved elemental chemical maps were built with selected X-ray line energy for each element: Bi (using M of 2430 eV ), Pt (using M of 2065 eV), and Pd (using L of 2850 eV).  To obtain good signals for the Bi, Pt, and Pd spectra, a dwell time of 10$\mu$s was used, resulting in a total acquisition time of about 4.6 min.  Parameters were optimized to minimize sample drift.

The magnetization measurements were done using Quantum Design MPMS3, equipped with iQuantum Helium3 cryostat that can measure in the 0.5-6~K range. The heat capacity measurements were performed with the PPMS equipped with a dilution refrigerator in the 0.05~K to 4~K temperature range.

The $\mu$SR measurements were carried out at TRIUMF's M15 beamline, equipped with a dilution refrigerator and 5~T superconducting magnet. For the zero field $\mu$SR measurements, the muons were implanted with their spins polarized antiparallel to the beam axis, and the field was zeroed to within two $\mu$T  following the procedure using muonium in GaAs at the sample position as described in Morris and Heffner \citep{Morris2003}. For transverse field measurements, the muon spins were rotated perpendicular to the beam axis before implantation.

\subsection{Structural Characterization}\label{sec:structural}
Bi$_2$PdPt crystallizes in a hexagonal crystal structure with space group $P6_3/mmc$.  By performing the Rietveld refinement of the powder XRD data using FullProf software, we obtained the lattice parameters  (a=b= 4.2819(3)~\AA, c= 5.5819(3) \AA,~$\alpha=\beta=90\degree, \gamma=120\degree$).  The Bi$_2$PdPt atomic arrangement was imaged along the c-axis, revealing the hexagonal arrangement of atoms expected for a hexagonal crystal structure. Interplanar distances measured from the STEM HAADF images agree with distances derived from the XRD analysis. The EDS analysis revealed the chemical nature and composition of each atomic column (Fig. 1b). Bi atoms occupy atomic sites, forming a typical honeycomb pattern, while Pd and Pt occupy sites of the same atomic column. \begin{figure*}[t]
\begin{center}
\includegraphics[width=\textwidth]{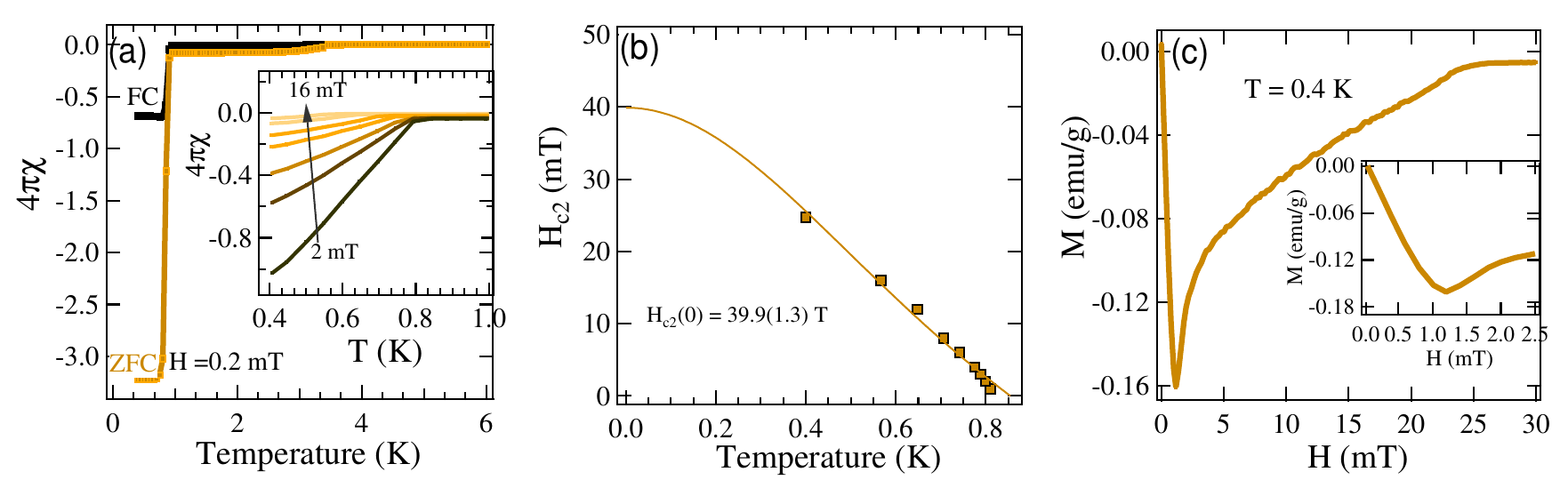}
\caption{\label{fig2} (a) Magnetic susceptibility measurement conducted in 0.2 mT field on Bi$_2$PdPt shows a bulk superconducting transition at $\approx$ 0.8 K for Bi$_2$PdPt. [Inset] Magnetic susceptibility as a function of temperature for different magnetic fields is shown here. (b)  The upper critical field, $H_{c2}$, as a function of temperature, fit to the Ginzburg-Laudau equation, is plotted. (c) The magnetization versus temperature curve suggests that the lower critical field, H$_{c1}$, is of the order of 1 mT at 0.4 K.  }
\end{center}
\end{figure*}
Furthermore, quantitative chemical analysis performed in regions of different sizes showed that the atomic compositions (to within an error of 3-5\%) for Bi, Pd, and Pt are 50\%, 25\%, and 25\%, respectively.

\subsection{Magnetization}
Magnetization measurements were carried out on a polycrystalline sample of Bi$_2$PdPt, which revealed the onset of bulk superconductivity at 0.8 K. The magnetization as a function of the field at 0.4 K [Fig. \ref{fig2}(c)] indicates that Bi$_2$PdPd is a type II superconductor, as it shows a vortex state beyond $\approx$ 1 mT field ($H_{c1}$).  We estimate the upper critical field, H$_{c2}(0)$,  to be 39.9(1.3) mT from the plot of the temperature dependence of magnetic susceptibility at different fields, as shown in Fig. \ref{fig2} (a-b).  Using $\mu_0 H_{c2}=\frac{\phi_0}{2\pi\xi^2}$, we obtained the value of the coherence length, $\xi$, to be 91(1)  nm.   In addition to the superconducting transition at 0.8 K, we observed a dip (less than 5\% of total susceptibility) in magnetic susceptibility at $\approx$ 3.2 K, which could be attributed to surface superconductivity or an unknown superconducting phase that could not be observed in XRD, specific heat or any of our microscopic measurements. Similar behavior  was observed in the SrPtAs \cite{srptas}.

\begin{figure}[]
\begin{center}
\includegraphics[width=0.47\textwidth]{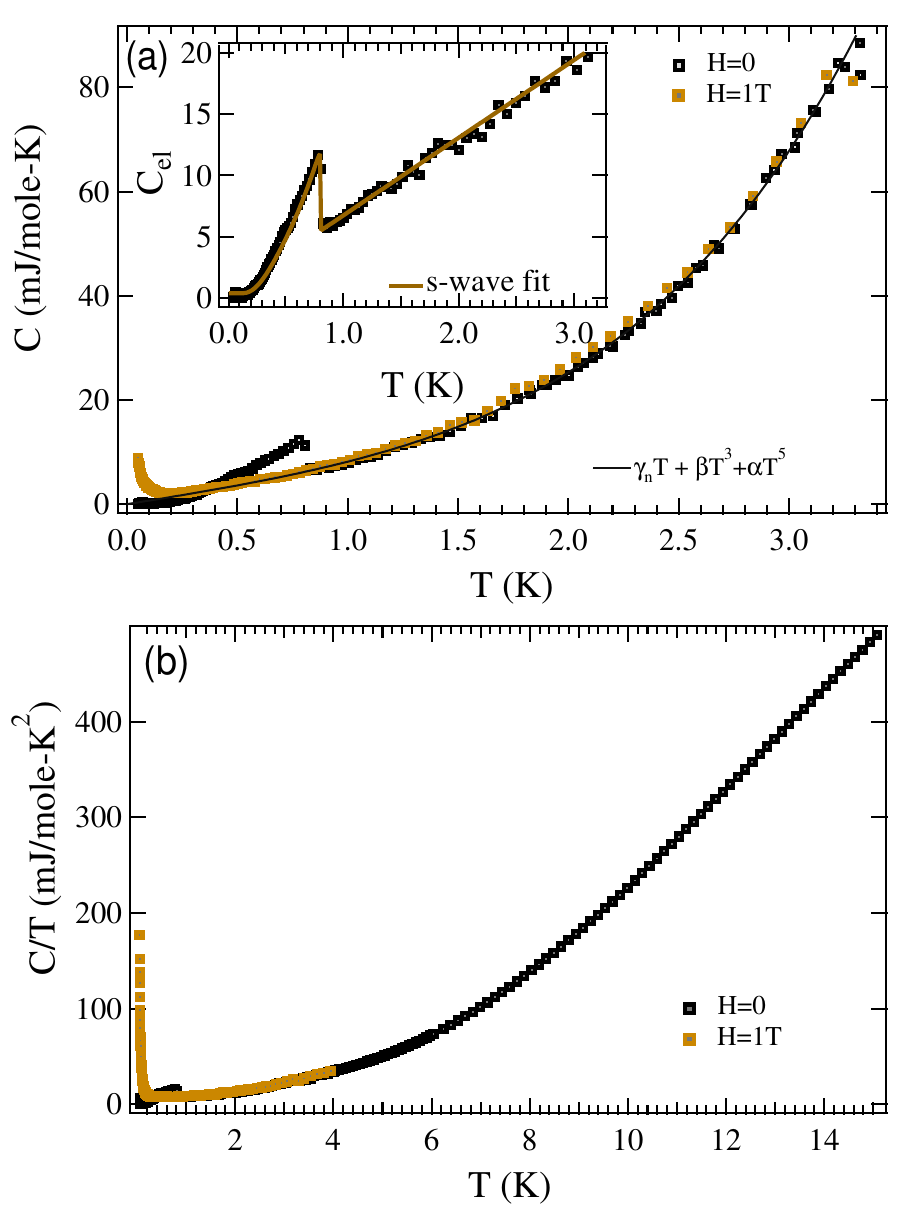}
\caption{\label{HeatCap} (a) Plot of heat capacity (C)  data in zero field shows a jump at 0.8 K corresponding to bulk superconducting transition, which disappears on application of 1 T field. The black line is the fit of the data to $\gamma_n T +\beta T^3 + \alpha T^5$, which allows us to extract the electronic part of the heat capacity.
 [Inset] The electronic heat capacity and its s-wave model fit is shown.   (b) The plot of zero and 1 T  field C/T data shows that there is only one superconducting transition in this sample, which occurs at 0.8 K. }
\end{center}
\end{figure}

\subsection{Specific Heat} Specific heat measurements performed on a polycrystalline sample of Bi$_2$PdPt show a sharp jump at $\approx$ 0.8 K [Fig. \ref{HeatCap}], consistent with the superconducting transition observed with magnetization measurements. 
The heat capacity in the normal state of a superconductor can be described as a sum of electronic and phononic terms such that $C =\gamma_n T +\beta T^3 + \alpha T^5$. We isolate the electronic portion of the specific heat by subtracting the phononic contribution ($\beta T^3+ \alpha T^5$). The values of the Sommerfeld coefficient ( $\gamma_n$ ),  $\beta$ and $\alpha$ are estimated from the fits are 6.34(4) mJ/mole-K$^2$, 1.39(5) mJ/mole-K$^4$, and 0.044(6) mJ/mole-K$^6$, respectively.   From the value of $\beta$, we can estimate the Debye temperature using the relation, $\theta_D = (12\pi^4N_Ark_B/5\beta)^{1/3} $, where r is the number of atoms per formula unit. The estimated $\theta_D$ is 112(1)~K.

One can fit the electronic heat capacity (C$_{el}$) and obtain important information about superconducting gap magnitude using the following  entropy (S) relation, 
\begin{equation}
    \frac{S}{\gamma_n T_c} = \frac{-6}{\pi^2k_B T_c}
\int_{0}^{\infty} [(1-f)\ln{(1-f)} + f\ln{f}]d\xi  
\end{equation}
where $f =(1+e^{E/k_BT})^{-1}$ such that fermion excitation energy is $E(\xi) = [\xi^2 + \Delta(T)^2]^{1/2} $,  where $\xi$ is the energy of electrons with respect to Fermi energy.    The temperature dependence of the superconducting gap, $\Delta(T)$ \cite{gapequation}, within BCS theory, is given by 
 \begin{equation}
\Delta(T) = \Delta(0)\tanh\left\{1.82\left(1.018\left(\frac{T_c}{T}-1\right)\right)^{0.51}\right\}
\label{gap}
\end{equation}
where $\Delta(0)$ is the gap value at 0 K.   The specific heat data for Bi$_2$PdPt fits well with an s-wave model with gap magnitude of  1.63(3), which is close to the BCS value of 1.73.  

\begin{figure*}[]
\begin{center}
\includegraphics[width=\textwidth]{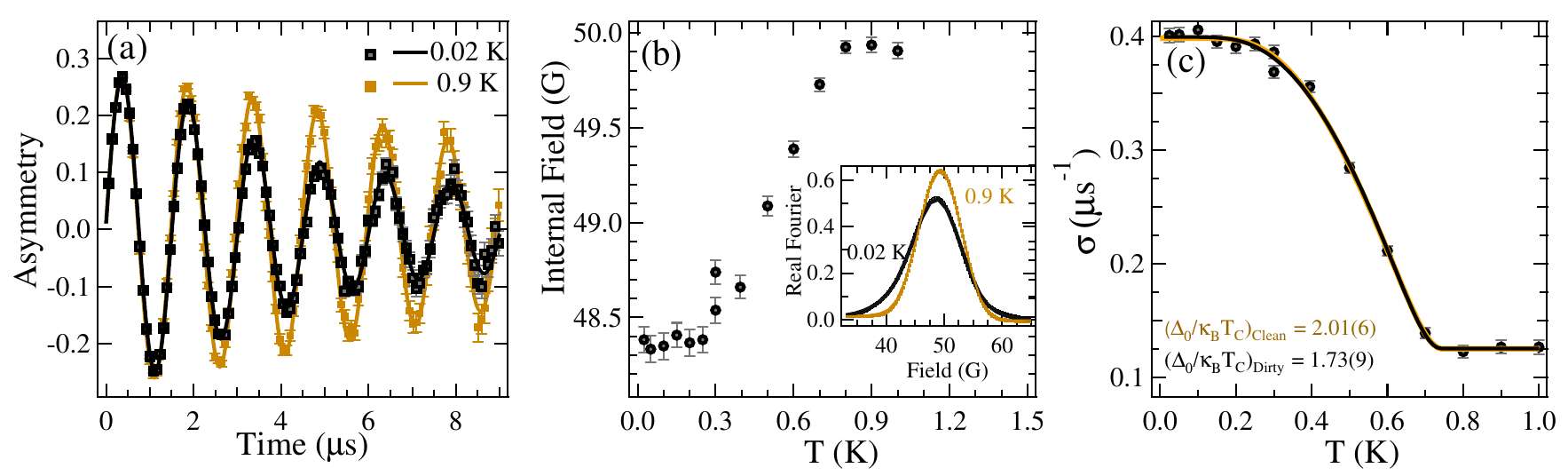}
\caption{\label{TF} (a) The asymmetry spectra, measured in 50 G field, collected above and below T$_c$,  show increased relaxation in the superconducting state. (b) The internal field of the sample was obtained from the fits of the asymmetry spectra. Bi$_2$PdPt shows zero Knight shift in the normal state, which starts decreasing below T$_c$.   (c) The plot of the relaxation rate parameter, $\sigma$, and its fit to the s-wave model in clean  and dirty limit are in good agreement, suggesting conventional s-wave superconductivity in Bi$_2$PdPt. }
\end{center}
\end{figure*}

\subsection{Muon Spin Relaxation and Rotation}
To further characterize the superconducting ground state of  Bi$_2$PdPt, we have utilized both muon spin relaxation and rotation measurements. Transverse field $\mu$SR can be utilized to study the vortex state of a type II superconductor and determine the temperature dependence of magnetic penetration depth $\lambda$(T). We cooled the samples in a 5 mT field during the transverse field $\mu$SR measurements, which is well above the lower critical field that ensures a well-ordered flux line lattice. The application of the field lowers the transition temperature of Bi$_2$PdPt to $\approx$ 0.75 K from its zero field value of 0.8 K.
In the mixed state of a superconductor, the flux line lattice (FLL) creates an inhomogeneous field distribution, which leads to the decay of the muon precession signal. The asymmetry spectra, as shown in Fig. \ref{TF}, were fit with a two-term oscillating decaying function. 
\begin{equation}
\begin{split}
G_{\mathrm{TF}}(t) = A[F \;\mathrm{exp}\left(-\sigma^2 t^2/2\right)\mathrm{cos}(\omega_{1}t+\phi)\\+\;(1-F)\;\mathrm{exp}\left(-\psi t\right)\mathrm{cos}(\omega_{2}t+\phi)]
\label{eqn3}
\end{split}
\end{equation}
 Here, the Gaussian decay term describes the sample signal, and the exponential decay term describes the temperature-independent background signal of the silver sample holder in the dilution refrigerator.  $\phi$ is the phase of the incident muons, whereas $\omega_1$ and $\omega_2$ are the frequency of the muon precession in the sample and background, respectively. F is the temperature-independent sample fraction, and $\sigma$ and $\psi$  are the sample and background relaxation rates, respectively. All the sample parameters, like $\sigma$ and $\omega_1$,  were kept temperature dependent, whereas the parameters describing the background, like $\omega_2$ and $\psi$,  were kept temperature independent while fitting the transverse field asymmetry spectra. 
 
 The total sample relaxation rate $\sigma$ has a temperature-dependent contribution from the FLL ($\sigma_{sc}$) and a temperature-independent contribution from nuclear moments ($\sigma_N$). The FLL contribution can be obtained via $\sigma_{sc} = \sqrt{\sigma^2-\sigma_N^2}$. The superconducting relaxation rate, $\sigma_{sc}$, describes the mean square inhomogeneity of the field ($\langle(\Delta B)^2\rangle$) due to the underlying FLL \cite{Aeppli1987}  seen by the muons, where $\langle(\Delta B)^2\rangle = \langle(B-\langle B\rangle)^2\rangle$. The superconducting relaxation rate is related to the field distribution through,
 \begin{equation}
 \label{variance}
     \sigma_{sc}^2 = \gamma_\mu^2\langle(\Delta B)^2\rangle  ,
 \end{equation}
 where $\gamma_\mu$ (= 2$\pi\times$135.5~MHz/T) is the muon gyromagnetic ratio. For small applied fields [H/H$_{c2} \ll$ 1], the superconducting relaxation rate can be used to calculate the superconducting penetration depth, $\lambda$, through Brandt's formula \cite{brandt} for a triangular Abrikosov vortex lattice:
 	\begin{equation}
    \sigma_{\mathrm{sc}}(T) = \frac{0.0609 \times \gamma_{\mu}\phi_{0}}{\lambda^{2}(T)}.
    \label{braneq2}
    \end{equation}
 Here, $\lambda(T)$ is in nm, $\sigma_{\mathrm{sc}}(T)$ is in $\mu$s$^{-1}$, and $ \phi_{0} $ (2.067$\times$10$^{-15}$ Wb) is the magnetic flux quantum. In the dirty limit, the superconducting gap and its temperature dependence, $\Delta(T)$, [equation \ref{gap}] can be calculated from the penetration depth via the following relation: 	\begin{equation}
 \frac{\lambda^{-2}(T)}{\lambda^{-2}(0)}  = \Biggl\langle\frac{\Delta(T)}{\Delta(0)}\mathrm{tanh}\left[\frac{\Delta(T)}{2k_{B}T}\right]\Biggr\rangle ,
\label{dirty s}
\end{equation} 
While in the clean limit, 
\begin{equation}
     \frac{\lambda^{-2}(T)}{\lambda^{-2}(0)} = 1+2\Biggl\langle\int_{|\Delta(T)|}^{\infty}\left(\frac{\delta f}{\delta E}\right)\frac{E dE}{\sqrt{E^{2}-\Delta^{2}(T)}}\Biggr\rangle  ,
    \label{eqnclean}
    \end{equation}  
 where $f = [1+\exp(E/k_{B}T)]^{-1}$ is the Fermi function, and the angular brackets denote the average over the Fermi surface.

From the fits of the TF-$\mu$SR data, we conclude that Bi$_2$PdPt hosts a conventional superconducting state with a normalized superconducting gap of 1.73(9) in the dirty limit, which is close to the BCS value of 1.732. Our analysis shows a s-wave gap of 2.01(6) in the clean limit. The comparison of the parameters obtained from the fits in the two limits is shown in Table I.  We would expect Bi$_2$PdPt to be in the dirty limit due to Pd/Pt site mixing; however, we can't provide conclusive evidence without resistivity data, which would have allowed for a determination of the mean free path ($l$).  If $\xi 
\ll l$,  the superconductor is in the clean limit; otherwise, if  $l 
\ll \xi$, the superconductor is in the dirty limit.

  \begin{table*}
   	\caption{Muon depolarization fitting parameters and resultant superconducting state parameters \newline(c.l. = clean limit, d.l.= dirty limit)}
   	\label{fit parameters}
   	\begin{center}
   		\begin{tabular*}{2.0\columnwidth}{l@{\extracolsep{\fill}}llllllll}\hline\hline
   			&Model& $\Delta_{0}$(meV) &$T_{c}$(K)&$\Delta_{0}/k_{B}T_{c}$ & $\chi^{2}$&$\lambda(0)(nm)$&$n_s/(m^*/m_e)$($10^{25} m^{-3}$)\\
   			\hline
   			\\[0.5ex] 
     		& s-wave d.l.& 0.111(6)&0.742(11)& 1.73(9)  & 1.05&532(2)&4.99(4)  \\
   			& s-wave c.l.& 0.129(3)&0.745(13)& 2.01(6) & 1.11&532(2) &4.99(4) \\
   			
   			\\[0.5ex]
   			\hline\hline
   		\end{tabular*}
   		\par\medskip\footnotesize
   	\end{center}
   \end{table*}

The zero temperature penetration depth, $\lambda(0)$, estimated through the clean s-wave model, is 532(2) nm. Subsequently, $\lambda(0)$ was used to calculate the \text{n$_s$/(m$^*/$m$_e$)} which comes out to be 9.96(7)$\times10^{25}$ $ m^{-3}$, where $n_s$ is the superfluid density and $m^*$ is the effective mass of the superconducting carrier. Using $m^*$=2, $n_s$ assumes a small value of 0.0176(1) per formulae unit. We estimate the Fermi temperature, $T_F$, to be 221 K, where we have used the Sommerfeld coefficient $\gamma_n$ = 6.34 mJ/K$^2$cm$^3$. The T$_c$/T$_F$ of $\approx$ 0.004 puts Bi$_2$PdPt in the vicinity of unconventional superconductors in the Uemura classification. According to Uemura, unconventional superconductors have $T_c/T_F$ between 0.01 and 0.1, and conventional superconductors have $T_c/T_F$ less than 0.001 \cite{Uemura1991, Uemura1989}. 



\begin{figure}[b]
\includegraphics[width=86.6mm]{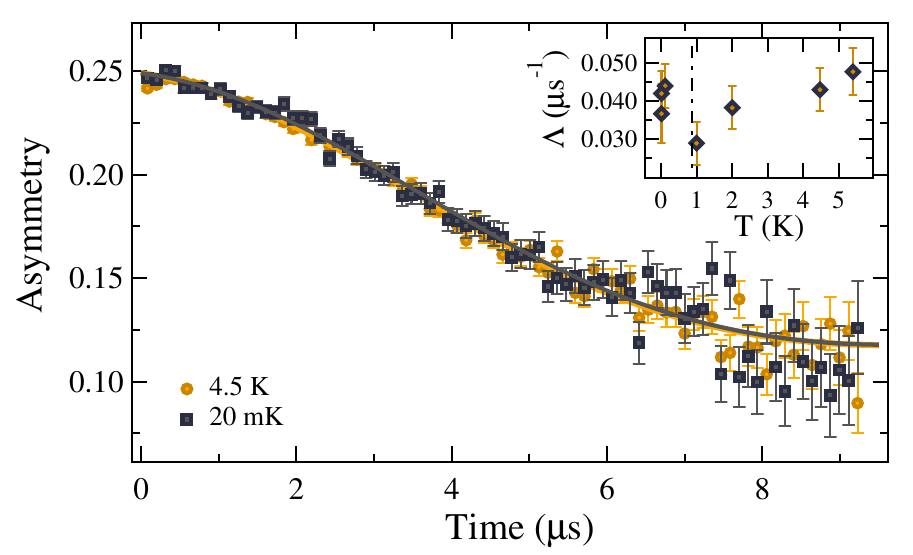}
\caption{\label{ZF} The zero field $\mu$SR spectra collected above and below the transition temperature of Bi$_2$PdPt. The spectra show no noticeable change in the relaxation rate, indicating the absence of spontaneous fields in the superconducting state. [Inset] The temperature dependence of $\Lambda$, representative of the internal field, doesn't show any noticeable increase onsetting at T$_c$ (0.8 K). }
\end{figure}

\paragraph*{Zero-field (ZF) muon spin relaxation/rotation ($\mu$SR)} 
Muon spin relaxation measurements performed in zero field (ZF) configuration can give clear evidence of spontaneous magnetic fields originating from the time-reversal symmetry (TRS) breaking superconducting state. We collected the ZF spectra for Bi$_2$PdPt on either side of the transition temperature, as shown in Fig. \ref{ZF}. The muon ensemble polarization decays due to random nuclear moments in the absence of static electronic or magnetic moments. This muon depolarization can be described by the Gaussian Kubo-Toyabe function G$_{KT}$(t) 
\begin{equation}
 G_{\mathrm{KT}}(t) = \frac{1}{3}+\frac{2}{3}(1-\sigma^{2}t^{2})\mathrm{exp}\left(-\frac{\sigma^{2}t^{2}}{2}\right) 
 \label{eqn1:zf}
 \end{equation} 
 where $ \sigma $ reflects the width of the nuclear dipolar field experienced by the muons and t is time.
 
 We use the following relaxation function to fit the ZF spectra
\begin{equation}
 A(t) = A_{1}G_{\mathrm{KT}}(t)\mathrm{exp}(-\Lambda t)+A_{\mathrm{BG}} 
 \label{eqn:tay}
 \end{equation}
 where $  A_{BG} $ is the background asymmetry, $  A_{1} $ is the sample asymmetry, and the term exp(-$ \Lambda $t) accounts for any additional relaxation channels (such as broken TRS). In a broken TRS state, a superconductor will exhibit an increased relaxation rate below the transition temperature \cite{sharma2023, luke1998}. However, for Bi$_2$PdPt, the spectra appear unchanged upon cooling through T$_c$, as shown in Fig. \ref{ZF}, and therefore, within our resolution, Bi$_2$PdPt preserves time-reversal symmetry. The inset of Fig. \ref{ZF} presents the relaxation rate parameter, $\Lambda$, obtained from the fits of the ZF data, varying randomly across the temperature points. In addition, we obtained a 0.179(3) $\mu s^{-1}$ for the Kubo-Toyabe relaxation rate parameter. From the $\Lambda$ plot, we estimate that 71 mG is the maximum possible time-reversal symmetry-breaking field that can go undetected in Bi$_2$PdPt.


\section{Density-Functional Theory Calculations}
\begin{figure*}[]
\begin{center}
    \includegraphics[width=\textwidth]{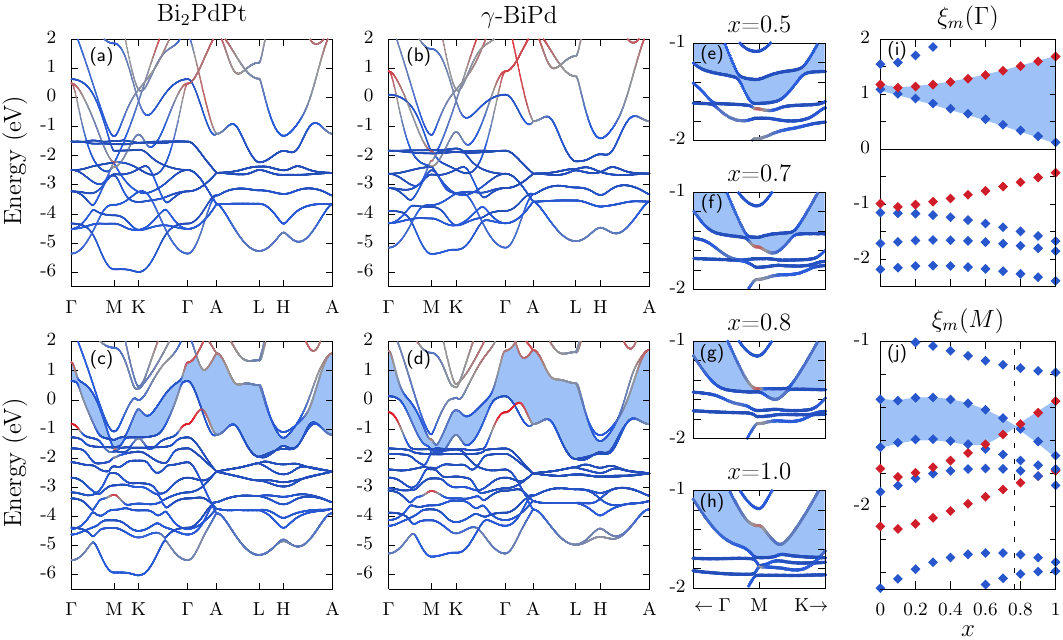}
    \vspace*{-0.5cm}
    \caption{DFT Band structure for Bi$_2$PdPt (a,c) and $\gamma$-BiPd (b,d) without SOC (a,b) and with SOC (c,d). The band character is colored by the projection onto Wannier orbitals centered on Bi atoms (red) and Pt/Pd atoms (blue). (e-h) Zoom of band structure near the M point for values of $x$ in the chemical formula, BiPd$_{x}$Pt$_{1-x}$. Band energies at the $\Gamma$ point (i) and $M$ point (j). The band energies in (i-j) are coloured according to even (blue) and odd (red) parity. The dashed line in (j) indicates the concentration at which the $Z_2$ index calculated for the highest 14 valence bands changes from 0 ($x<0.75$) to 1 ($x>0.75$). The blue shaded region in (c-j) covers the gap separating the 14th and 15th highest valence bands.}
    \label{fig:wannier}
\end{center}
\end{figure*}

\subsection{Electronic Band Structure}
We have performed fully relativistic and scalar relativistic density-functional theory calculations in the plane-wave-pseudopotenial formalism as implemented in Quantum Espresso~\cite{qe1,qe2}. The generalized gradient approximation~\cite{PBE} was used to approximate the exchange-correlation potential. Ultrasoft pseudopotentials as provided in the PSlibrary associated with Quantum espresso were used. Cold smearing~\cite{Marzari_Payne_1999} was included in the DFT calculation with a smearing parameter of 0.01 Ry. A 12$\times$12$\times$12 Monkhorst-Pack $\vec{k}$-point mesh is used to sample the Brillouin zone. Kinetic energy cutoffs of 80 Ry for the wavefunctions and 640 Ry for the charge density were used. We use the virtual-crystal approximation to model disorder by constructing a pseudopotential for a virtual Pd$_{x}$Pt$_{1-x}$ ($0\leq x \leq 1$) atom from the pseudopotentials of Pd and Pt as implemented in the virtual\_v2.x program within Quantum Espresso. Each system considered is relaxed until pressures in every direction are less than 0.05 kbar. The resulting lattice constants for Bi$_2$PdPt are $a=b=4.42$~\AA\ and $c=5.64$~\AA, in reasonable agreement with the experimental results from section~II\ref{sec:structural}. We find through orbital projections of the calculated Kohn-Sham states that the Bi-p orbitals are strongly hybridized with the Pd/Pt-s, Pd/Pt-p, and Pd/Pt-d orbitals in a wide energy window of approximately [-6,6]~eV. Wannier90~\cite{w90} was used to calculate maximally-localized Wannier functions (MLWF). Bi-s, Bi-p, Pd/Pt-s, Pd/Pt-p, and Pd/Pt-d projected orbitals were used to initialize the wannierization. A frozen disentanglement window of [-6,4]~eV was chosen to capture the majority of the Bi-p and Pd/Pt-d states.

The DFT band structure near the Fermi energy and band character can be seen in Fig.~\ref{fig:wannier}. The bands are of primarily Pd/Pt character, with two bands of primarily Bi character near the $\Gamma$ point. We have found that the inclusion of spin-orbit coupling produces band inversion at the $\Gamma$ point in the 14th valence band around [0,1.5]~eV, which is reflected in the change in band character at this point. A continuous direct gap can be seen in the band plot separating the 14th and 15th valence bands when SOC is included. For $\gamma$-BiPd, We have calculated the energy difference between the 14th and 15th band across dense grids of 24$^3$, 32$^3$, and 40$^3$ evenly spaced $\vec{k}$-points in the Brillouin zone and have found a minimum band gap of 106 meV, confirming the presence of a band gap throughout the Brillouin zone.

\subsection{Topological Classification}
Since Bi$_2$PdPt has a centrosymmetric crystal structure, the $Z_2$ indices, denoted by $(\nu_0;\nu_1,\nu_2,\nu_3)$, can be calculated from parity products evaluated at eight time-reversal invariant momenta (TRIM)~\cite{Fu_Kane_2007}. We denote parity products at the TRIM, $\frac{n_1}{2}\vec{b}_1+\frac{n_2}{2}\vec{b}_2+\frac{n_3}{2}\vec{b}_3$, by $\delta_{(n_1,n_2,n_3)} = \prod_{m}\xi_m(n_1,n_2,n_3)$, where the $\xi_m$ are parity eigenvalues and the product is taken over bands below a band gap. The strong $Z_2$ index, $\nu_0$, is obtained from $(-1)^{\nu_0} = \prod_{n_1,n_2,n_3}\delta_{(n_1,n_2,n_3)}$, and the weak $Z_2$ indices, $\nu_{i=1,2,3}$, are obtained from $(-1)^{\nu_i} = \prod_{n_i=1,n_j,n_k}\delta_{(n_1,n_2,n_3)}$ ($j\neq k \neq i$). The calculated parities and parity products of the highest 14 Kramers-degenerate (28 in total) valence bands for Bi$_2$PdPt and $\gamma$-BiPd are given in Table~\ref{tab:parities}. The symmetry-enforced degeneracy of bands at the $A$ and $L$ points originating from the non-symmorphic lattice structure cause four-fold degeneracy of each band consisting of two Kramers pairs of opposite parity denoted by "(+,$-$)" in Table~\ref{tab:parities}. All weak $Z_2$ indices are therefore trivial, and the strong index is determined from $(-1)^{\nu_0} = \delta_{\Gamma}\delta_{M}$. We find that the calculated $Z_2$ indices, $(\nu_0;\nu_1,\nu_2,\nu_3)$, are $(1;0,0,0)$ for $\gamma$-BiPd and $(0;0,0,0)$ for Bi$_2$PdPt. 

\begin{table}[h!]
  \centering
  \begin{tabular}{| c |c c c c c c c c c c c c c c | c |} 
    \hline
    \multicolumn{16}{|c|}{$\gamma$-BiPd ($\nu_0=1$)} \\
    \hline
    & 1 & 2 & 3 & 4 & 5 & 6 & 7 & 8 & 9 & 10 & 11 & 12 & 13 & 14 & $\delta$ \\
    \hline
    $\Gamma$ & + & $-$ & + & + & + & + & + & + & + & + & + & + & $-$ & + & + \\
    \hline
    $M$ & $-$ & + & + & + & + & + & $-$ & + & + & + & + & $-$ & + & + & $-$ \\
    \hline
    $L$ & (+ & $-$) & (+ & $-$) & (+ & $-$) & (+ & $-$) & (+ & $-$) & (+ & $-$) & (+ & $-$) & $-$ \\
    \hline
    $A$ & (+ & $-$) & (+ & $-$) & (+ & $-$) & (+ & $-$) & (+ & $-$) & (+ & $-$) & (+ & $-$) & $-$ \\
    \hline
  \end{tabular}
  \\ \phantom{A} \\
  \begin{tabular}{| c |c c c c c c c c c c c c c c | c |} 
    \hline
    \multicolumn{16}{|c|}{Bi$_2$PdPt ($\nu_0=0$)} \\
    \hline
    & 1 & 2 & 3 & 4 & 5 & 6 & 7 & 8 & 9 & 10 & 11 & 12 & 13 & 14 & $\delta$ \\
    \hline
    $\Gamma$ & + & $-$ & + & + & + & + & + & + & + & + & + & + & $-$ & + & + \\
    \hline
    $M$ & $-$ & + & + & + & + & + & $-$ & + & + & + & $-$ & + & $-$ & + & + \\
    \hline
    $L$ & (+ & $-$) & (+ & $-$) & (+ & $-$) & (+ & $-$) & (+ & $-$) & (+ & $-$) & (+ & $-$) & $-$ \\
    \hline
    $A$ & (+ & $-$) & (+ & $-$) & (+ & $-$) & (+ & $-$) & (+ & $-$) & (+ & $-$) & (+ & $-$) & $-$ \\
    \hline
  \end{tabular}
  \caption{Parities and parity products of the highest 14 valence bands for $\gamma$-BiPd and Bi$_2$PdPt.}
  \label{tab:parities}
\end{table}

Given the difference in strong $Z_2$ index between Bi$_2$PdPt ($\nu_0=0$) and $\gamma$-BiPd ($\nu_0=1$), it is expected that a topological phase transition occurs in the more general system BiPd$_{x}$Pt$_{1-x}$ for a value of $x$ between 0.5 and 1. As shown in Fig.~\ref{fig:wannier}. At approximately $x=0.75$, the direct gap between the 14th and 15th valence bands closes, and the parity of the 14th band changes from $+1$ to $-1$ as $x$ is decreased from the topologically non-trivial state at $x=1$. Our DFT calculations, therefore, indicate that BiPd$_{x}$Pt$_{1-x}$ is potentially a topological superconductor for substitution fractions of $x>0.75$.

\begin{figure}[]
\begin{center}
    \vspace*{0.11cm}
    \includegraphics[width=\columnwidth]{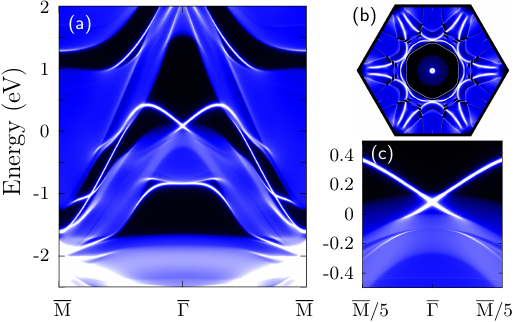}
    \caption{Surface state spectral function for $\gamma$-BiPd with a Pd-terminated (001) surface. (a) Energy dependence of the surface states for a path between the TRIM $\overline{\Gamma}$ and $\overline{\rm M}$ points. (b) Surface-state Fermi surface. (c) Same as (a) zoomed in near the Dirac surface state at the $\overline{\Gamma}$ point.}
   \label{fig:SurfaceStates}
\end{center}
\end{figure}

In light of the non-trivial band topology determined for $\gamma$-BiPd, we have use the tight-binding Hamiltonian in the MLWF basis for $\gamma$-BiPd to calculate the surface-state spectral function for a semi-infinite slab using the iterative method given in Ref.~\cite{Sancho_Rubio_1985} as implemented in the WannierTools program~\cite{wanniertools}. The resulting surface states for the Pd-terminated (001) surface are shown in Fig.~\ref{fig:SurfaceStates}. A Dirac cone can be seen in the surface state spectral function at the $\Gamma$ point within approximately 0.1~eV of the Fermi energy and near -1.5~eV at the $M$ points where Kramers degeneracy is enforced.

\section{Discussion}
Arguments made by Fu and Kane~\cite{Fu_Kane_2008} suggest that topologically-protected Dirac surface states in proximity to s-wave superconductivity have the potential to realize topological superconductivity characterized by Majorana fermion bound states localized within vortex cores. This scenario naturally occurs in materials with intrinsic s-wave bulk superconductivity and non-trivial band topology through the bulk-boundary correspondence. For example, DFT-based calculations indicate that this may be the case for FeSe$_{0.5}$Te$_{0.5}$~\cite{Wang_Fang_2015} and $\beta$-Bi$_2$Pd~\cite{Benia_Wahl_2016}. The calculations of Ref.\cite{Wang_Fang_2015} and Ref.\cite{Benia_Wahl_2016} have shown band inversion originating from strong spin-orbit coupling leads to non-trivial band topology and Dirac surface states in FeSe$_{0.5}$Te$_{0.5}$ and $\beta$-Bi$_2$Pd respectively. Additionally, calculations within Eliashberg theory for $\beta$-Bi$_2$Pd agree well with experimental measurements, supporting the existence of a fully-gapped superconducting state arising from an electron-phonon pairing mechanism~\cite{Zheng_Margine_2017}.

Bi$_2$PdPt, unlike earlier report \cite{anshu}, displays a bulk superconducting transition at 0.8 K. To demonstrate the phase purity and uniformity of our polycrystalline samples, we have employed XRD and microscopic probes like HAADF-STEM. We characterized the superconducting properties using magnetization, heat capacity, and $\mu$SR. The heat capacity data suggests conventional isotropic gapped superconductivity.  The $\mu$SR measurements also present evidence for the s-wave type superconductivity. In addition, the zero field $\mu$SR exhibits a time-reversal preserving superconducting state consistent with BCS theory. If the concentration $x$ is varied in our DFT calculations, we find that there is a critical concentration $x_c\approx 0.75$ in the isostructural compounds BiPd$_x$Pt$_{1-x}$, at which a topological phase transition occurs. Our calculations, therefore, indicate that Bi$_2$PdPt is topologically trivial. On the other hand, the DFT results suggest that a topological superconducting state may be realized in BiPd$_x$Pt$_{1-x}$ as the concentration of Pd is increased, assuming s-wave superconductivity is maintained. Our calculations for the topologically non-trivial $\gamma$-BiPd ($x$=1), show that the band gap separating inverted bulk bands is closed by a Dirac surface state close to the Fermi energy. The Dirac states that we find in our DFT calculations may be compared with Dirac surface states imaged in ARPES experiments of other Bi- and Pd-based compounds. For example, Dirac surface states with binding energies of about 0.7~eV, 1.26~eV, and 2.41~eV have been found in $\alpha$-BiPd~\cite{Benia_Wahl_2016}, $\alpha$-Bi$_2$Pd~\cite{Dimitri_Neupane_2018}, and $\beta$-Bi$_2$Pd~\cite{Sakano2015} respectively, and we reiterate that STM measurements on $\beta$-Bi$_2$Pd thin films show promising evidence for Majorana excitations associated with topological superconductivity~\cite{Lv_Xue_2017}.

\section{Conclusion}

We have characterized the superconducting state of Bi$_2$PdPt through model fits to $\mu$SR data in the superconducting state, and the band topology of the more general BiPd$_x$Pt$_{1-x}$ system, of which Bi$_2$PdPt is a special case with $x$=$0.5$. Our calculations suggest that Bi$_2$PdPt is a topologically trivial material and that a topological phase transition from a trivial $Z_2$ topological metal to a non-trivial $Z_2$ topological metal may be induced by increasing the concentration of Pd. While BiPd typically forms in the non-centrosymmetric $\alpha$ phase at low temperatures, it is clear that Bi$_2$PdPt is stable in the hexagonal phase, and we find it reasonable that the hexagonal phase may be stabilized for concentrations of $x>0.5$, of which $\gamma$-BiPd is a limiting case ($x=1$). Our results suggest bulk superconductivity in Bi$_2$PdPt with $T_c$ (0.8 K), which is lower than the previous report \cite{anshu}. We report the results of XRD, HAADF-STEM, EDS, magnetization, heat capacity, and $\mu$SR measurements to characterize the bulk and microscopic properties in normal and superconducting states.  The transverse field $\mu$SR measurements suggest s-wave superconductivity in Bi$_2$PdPt. Our zero field $\mu$SR measurements display no change in $\mu$SR relaxation rates below T$_c$, indicating a time reversal preserving superconducting state. Our results motivate further experimental and theoretical exploration of the possibility of topological superconductivity in the BiPd$_x$Pt$_{1-x}$ series.

\section{Acknowledgments}
Work at McMaster was supported by the Natural Sciences and Engineering Research of Council of Canada. R. P. S. acknowledges the Science and Engineering Research Board, Government of India, for the Core Research Grant CRG/2019/001028. The financial support from DSTFIST Project No. SR/FST/PSI-195/2014(C) is also thankfully acknowledged. This research was enabled in part by support provided by SHARCNET (sharcnet.ca) and the Digital Research Alliance of Canada (alliancecan.ca). M.J.L. and B.S.A. acknowledge the financial support of the Natural Sciences and Engineering Research Council of Canada (NSERC) under the Discovery Grant Program. We also thank the Canadian Centre for Electron Microscopy (CCEM) for providing access to electron microscopy facilities.

\bibliography{Bi2PdPt}

\end{document}